\documentclass[fleqn,twoside]{article}
\usepackage{espcrc2}
\usepackage{graphicx}
\usepackage{amsmath}
\usepackage{amssymb}

\title{
$\check{C}$erenkov angle and charge reconstruction with the 
RICH detector of the AMS experiment
} 

\author{F.Barao\address[lip]{LIP-Av. Elias Garcia, 14, 1000-Lisboa, Portugal}
               \address[ist]{IST-Av. Rovisco Pais, 1000-Lisboa, Portugal},
        L.Arruda\addressmark[lip],
        J.Borges\addressmark[lip],
        P.Gon\c{c}alves\addressmark[lip],
        M.Pimenta\addressmark[lip]\addressmark[ist],
        I.Perez\address[fcul]{FCUL, Campo Grande, 1749-Lisboa}
       }

\begin{document}

\begin{abstract}
The Alpha Magnetic Spectrometer (AMS) experiment to be installed 
on the International Space Station (ISS) will be equipped with a 
proximity focusing Ring Imaging $\check{C}$erenkov (RICH) detector,
for measurements of particle electric charge and velocity.
In this note, two possible methods for reconstructing the $\check{C}$erenkov
angle and the electric charge with the RICH, are discussed.
A Likelihood method for the $\check{C}$erenkov angle reconstruction was 
applied leading to a velocity 
determination for protons with a resolution of around 0.1\%.
The existence of a large fraction of background photons which can
vary from event to event, implied a charge reconstruction method
based on an overall efficiency estimation on an event-by-event basis.
\end{abstract}

\maketitle

\section{The AMS experiment}
The Alpha Magnetic Spectrometer (AMS) \cite{bib:ams} is a precision spectrometer to be 
installed by 2005 in the International Space Station (ISS), where it will
operate for a period of three years.
Its main goals are the search for cosmic anti-matter, the search for dark matter 
and the measurement of the relative abundance of elements and isotopes in
primary cosmic rays.

The future installation of AMS in the ISS was preceded by a 10 days 
engineering test flight aboard the Space Shuttle Discovery in June 1998,
at a mean altitude of 370 km. 
Although the purpose of this experimental flight was the test of the 
spectrometer design principles,
about 100 million events were
collected enabling precise measurements of the spectra of high energy protons, 
electrons, positrons and  helium nuclei \cite{bib:ams-publications}. 

\begin{figure}[htb]
\begin{center}
\scalebox{0.36}{%
\includegraphics[bb=14 14 549 546]{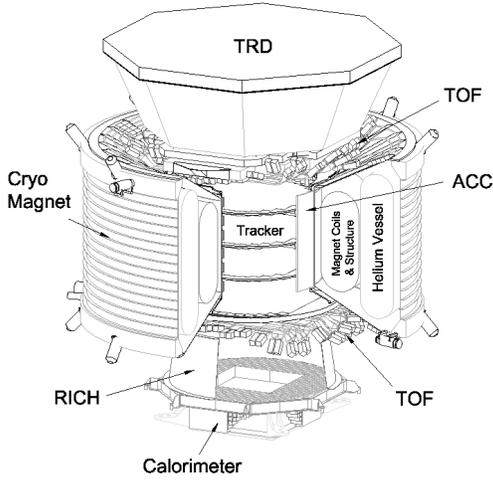}
} 
\caption{A whole view of the AMS Spectrometer. 
}
\end{center}  
\label{fig:ams}
\end{figure}

The AMS spectrometer capabilities were extended 
with respect to those of the experimental flight, 
through the inclusion of new subdetector systems and the completion of others. 
The spectrometer design includes a superconducting magnet, a Time-of-Flight system (TOF), a Silicon Tracker, Veto Counters,
a Transition Radiation Detector (TRD), an Electromagnetic Calorimeter (ECAL) 
and a Ring Imaging $\check{C}$erenkov Detector (RICH). 
It will be 
capable of measuring the ri\-gi\-di\-ty ($R\equiv pc/ |Z| e$), the charge ($Z$),
the velocity ($\beta$) and the energy ($E$) of cosmic rays within a 
geometrical acceptance of $ \sim 0.5~m^2.sr$.
The tracking system, with a cylindrical shape, 
is made of 8 double sided silicon planes embedded inside a magnetic 
field of about $0.9$ Tesla and 
will provide both charge measurements and momentum measurements with a 
resolution $\Delta p/p$ of at most 2\% up to 100 GeV/c/nucleon.
The TOF system, made of four scintillator planes placed at the magnet 
end-caps, will provide 
a fast trigger, charge and velocity 
measurements for charged particles as well as information on their 
incidence direction.
On the top of the spectrometer, the TRD will discriminate between 
leptons and hadrons, and, on the bottom, the ECAL will contribute 
to the $e/p$ separation and will measure the energy of the gamma rays crossing AMS.
Figure \ref{fig:ams} shows a schematic view of the AMS spectrometer.

\begin{figure}[htb]
\begin{center}
\scalebox{0.28}{%
\includegraphics[bb=32 22 830 495]{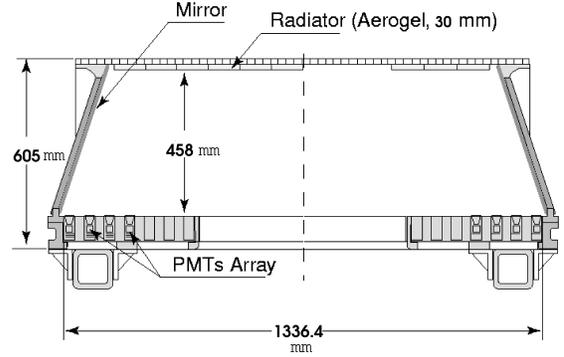}
} 
\caption{The RICH detector.\label{fig:rich}}
\end{center} 
\end{figure}

The RICH detector will operate between the TOF and the ECAL subdetectors.
It was designed to measure the velocity of singly charged particles with
a resolution $\Delta \beta/ \beta$ of 0.1\%, to extend the electric charge 
separation up to the Iron element and to contribute to the albedo rejection.
Its acceptance is of $\sim 0.4 ~m^2.sr$, that is, around 80\% of the 
AMS acceptance.
The RICH is a pro\-xi\-mi\-ty focusing detector with
a low refractive index radiator (aerogel) on the 
top and a pixelized photomultiplier matrix 
on the bottom, where the radiated $\check{C}$erenkov photons are collected.
The active pixel size is of $8.5~mm$.
A conical shaped mirror surrounds the whole set, increasing the detector 
reconstruction efficiency.
Constraints on the amount of heterogeneous matter 
in front of the downstream electromagnetic calorimeter have imposed 
a large non-active (empty) readout area in the detection plane.
For a more detailed  des\-cription of the detector see reference 
\cite{bib:buenerd} in these proceedings.
Figure \ref{fig:rich} shows a view of the RICH detector and of its dimensions.

\section{Velocity reconstruction\label{sec:velocity}}
A charged particle 
crossing the RICH radiator material of refractive index $n$,
emits photons if its velocity ($\beta$) is 
larger than the velocity of light in that medium.
The aperture angle of the emitted photons with respect to the 
radiating particle is known as the $\check{C}$erenkov angle, $\theta_c$,
and it is given by \cite{bib:rich}:
\begin{equation}
\cos\theta_c = \frac{1}{\beta~n}
\label{eq1}
\end{equation}
It follows that the velocity of the particle, $\beta$, is straightforwardly 
derived from the $\check{C}$erenkov angle reconstruction.

\begin{figure}[htb]
\begin{center}
\scalebox{0.4}{%
\includegraphics[bb=20 22 521 407]{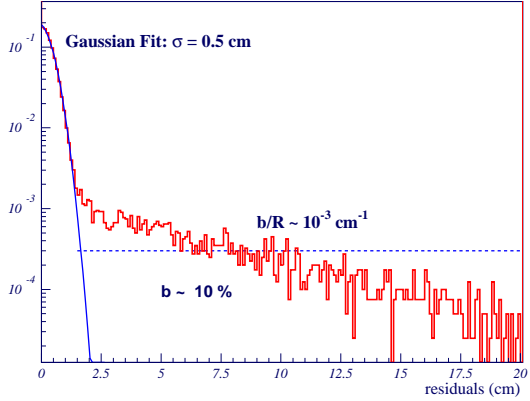}  
}                                                           
\caption{Distribution of the hit residuals 
with respect to the expected $\check{C}$erenkov pattern.
\label{fig:residuals}}
\end{center}  
\end{figure}

The emitted photons 
can suffer interactions in the aerogel radiator (Rayleigh scattering, absorption),
can be reflected or absorbed on the mirror surface and 
can fall on the active area composed of solid light guides on 
top of the photomultipliers.
As a consequence, 
the reconstruction of the $\check{C}$erenkov angle has 
to deal with two kinds of photons; 
those which are only slightly deviated from the expected photon pattern
due to the pixel granularity, radiator thickness and chromaticity effects, 
and,
those which spread all over the detector, 
faked by photomultipliers noise and due to photon scattering.
The former, cor\-res\-pon\-ding to the Signal, produce the $\check{C}$erenkov 
photon pattern. They are Gaussian distributed, with a width $\sigma\sim 0.5~cm$, 
reflecting essentially the uncertainty related to the pixel size. 
The latter, constitute an essentially flat background modulated by the geometry 
of the detection plane.
Figure \ref{fig:residuals} shows the distribution of the hit residuals 
with res\-pect to the expected $\check{C}$erenkov pattern for a 3~cm thick,
1.03 refractive index aerogel radiator.
The probability density function for a detected hit 
to belong to the pattern is therefore expressed as:
\begin{equation}
p = (1-b) \frac{1}{\sigma\sqrt{2\pi}} 
    \exp{ \left[ -\frac{1}{2}\left( \frac{r_i}{\sigma} \right)^2 \right] } + 
\frac{b}{d} 
\end{equation}
where $b$ is the photon background fraction, $b/d$ is the background fraction per 
unit of distance ($\sim10^{-3}/cm$) and 
$r_i$ is the closest distance from the hit $i$ to the photon pattern.

\begin{figure}[htb]
\begin{center}
\scalebox{0.35}{%
\includegraphics[bb=55 155 520 625]{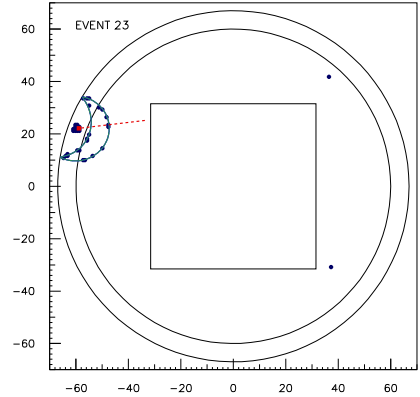}  
}                                                           
\caption{Reconstruction of a simulated helium event. 
The reconstructed photon pattern (full line) includes both, reflected and non-reflected
branches. The inner and outer circular lines 
correspond respectively to the upper and lower boundaries of 
the conical mirror. The square is the limit of the non-active region.\label{fig:event}}
\end{center}  
\end{figure}

\begin{figure}[htb]
\begin{center}
\scalebox{0.40}{                                                             
\includegraphics[bb=20 10 515 403]{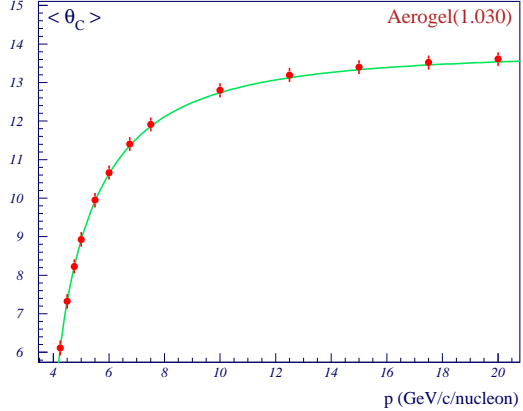}
}
\caption{$\check{C}$erenkov angle reconstructed as function of particle momentum 
         for an aerogel radiator.
\label{fig:thetac}        }
\end{center}  
\end{figure}

Complex photon patterns can occur at the detector plane 
due to mirror reflected photons, as can be seen on figure \ref{fig:event}. 
The $\check{C}$erenkov angle reconstruction procedure 
relies on the information of the particle direction 
provided by the Tracker, 
which, once extrapolated to the
radiator, provides an estimation of the mean photon emission vertex.
The tagging of the hits signalling the passage 
of the particle through the solid light guides in the detection plane, 
provides an additional track ele\-ment.
This, can be used as an additional track selection criterion. 
The best value of $\theta_c$ will result from the maximization of a Likelihood function, 
built as the product of the probabilities that the detected hits belong to a given (hypothesis)
$\check{C}$erenkov photon pattern ring,
\begin{equation}
L(\theta_c) = \prod_{i=1}^{nhits} p_i  \left[ r_i(\theta_c) \right].
\label{eq:likelihood}
\end{equation}
The RICH setup was fully simulated with GEANT3~\cite{bib:GEANT3} for radiators with 
different thickness and refractive index.
In order to trust the reconstruction, 
only events with at least 3 hits close (within $\sim$ 1.5cm) 
to the reconstructed photon pattern were selected.
Figure~\ref{fig:thetac} shows the reconstructed $\check{C}$erenkov angle for 
simulated protons with various momenta.
The $\check{C}$erenkov angle single hit resolution, obtained with aerogel of 1.03 refractive index
and 3 cm thick, was $\Delta \theta_c \sim 5~mrad$. 
This resulted in a velocity resolution $\Delta\beta/\beta$ slightly better 
than 0.1\% for protons ($\beta\sim 1$) and better than 0.05\% for helium nuclei.

\section{Charge reconstruction}
The $\check{C}$erenkov photons are uniformly emitted along the particle path 
inside the dielectric medium, $L$, and their number per unit of energy 
depends on the particle's charge, $Z$, and velocity, $\beta$, and on 
the refractive index, $n$, according to the expression \cite{bib:rich}:
\begin{equation}
\frac{dN_{\gamma}}{dE} \propto Z^2 L \left( 1-\frac{1}{\beta^2 n^2} \right)
\label{eq:dnde}
\end{equation}
Various factors contribute to the loss of some of these 
photons in the RICH;
radiator interactions ($\varepsilon_{rad}$), 
geometrical acceptance ($\varepsilon_{geo}$),
light guide efficiency ($\varepsilon_{lg}$) 
and photomultiplier quantum efficiency ($\varepsilon_{pmt}$).
Accordingly, the number of counted photoelectrons in the detector is given by:
\begin{equation}
n_{p.e.} \sim N_{\gamma}~\varepsilon_{rad}~\varepsilon_{geo}~\varepsilon_{lg}~\varepsilon_{pmt}  
\end{equation}

\begin{figure}[htb]
\scalebox{0.4}{%
\includegraphics[bb=3 2 545 403]{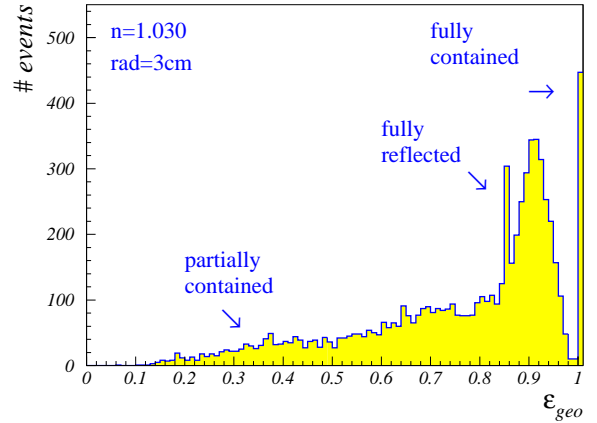}  
}                                                           
\caption{Photon geometrical acceptance for events in the RICH detector.\label{fig:geomacc}}
\end{figure}

All the efficiency factors, but the PMT efficiency, 
depend on the particle direction and of its incidence point on the radiator.
The radiator factor depends on the distance, $d$,
traversed by the photons inside the radiator. It is calculated by 
integrating the probability of a photon not to interact 
in the radiator, $\bar{p}_{\gamma} = e^{-d(z,\varphi)/L_{int}}$, 
along the radiator thickness and along the photon azimuthal angle ($\varphi$).
The geometrical acceptance accounts for photons lost 
through the radiator walls or totally reflected on media transitions, 
photons absorbed by the mirror and 
photons falling into the non-active detection area.
It is calculated taking into account the portion of the visible photon pattern
in units of photon azi\-mu\-thal angle, $\varepsilon_{geo} = \Delta\varphi/2\pi$.
Figure \ref{fig:geomacc} shows the geometrical acceptance 
calculated for an aerogel radiator of 1.030 and 
for events within the AMS fiducial volume.
The extreme variation of $\varepsilon_{geo}$ from event to event is clear.
The light guide efficiency factor depends on the incidence 
angle of the photons ($\theta_\gamma$) on the top of the light guide. 
It is calculated using the probability of a given 
photon to get to the photomultiplier cathode once it entered the light guide,
and by integrating it along the reconstructed photon pattern.

The charge of the radiating particles is derived 
from the number of photoelectrons, $n_{p.e.}$, close to the 
previously reconstructed photon pattern (see section \ref{sec:velocity}).
The number of radiated photons is obtained by correcting 
$n_{p.e.}$ by the overall event efficiency, which can be written as:
\begin{align}
\nonumber
& \varepsilon_{tot}= \frac{1}{2 \pi H_{rad}} \int_0^{H_{rad}} dz \sum_i^{n_{paths}} \rho_i  \\
& \int_{\varphi^{min}_i}^{\varphi^{max}_i} ~d\varphi \biggl [ e^{-\frac{d(z,\varphi)}{L_{int}}} 
~\varepsilon_{lg} (\theta_\gamma)
~\varepsilon_{pmt} \biggr ]
\end{align}
where $H_{rad}$ is the radiator thickness, 
$n_{paths}$ is 
the number of visible branches constituting the reconstructed pattern 
(i.e. reflected and direct branches) and $\rho_i$ being the reflectivity for the $i^{th}$ path.
 
\begin{figure}[htb]
\scalebox{0.4}{                                                             
\includegraphics[bb=3 2 545 403]{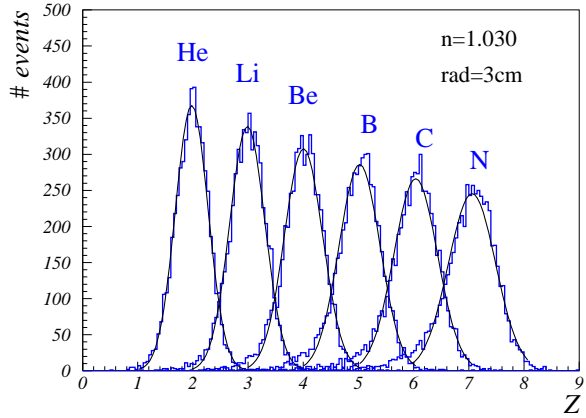}  
}
\caption{Charge reconstruction with simulated data in the RICH detector. 
Gaussian fits are superimposed to the distributions.\label{fig:charge}}
\end{figure}

The charge is then calculated according to expression \ref{eq:dnde}, 
where the normalisation constant 
can be evaluated from a calibrated beam of charged particles.
In the case of the present results it was obtained from 10000 simulated 
helium nuclei events.
Figure \ref{fig:charge} shows the reconstructed charge for ele\-ments ranging from 
Helium to Nitrogen. 
It was obtained using a 3 cm thick, 1.03 refractive index aerogel radiator,
for events with geometrical acceptance greater than 60\%.
The charge resolution ranges from $\Delta Z/Z \sim 15\%$ for Helium to 
$\Delta Z/Z \sim 5.5\%$ for Nitrogen.

\section{Conclusions}
AMS is a spectrometer designed for anti-matter, dark matter searches and for mea\-su\-ring
relative abundances of nuclei and isotopes.
Its installation in the International Space Station is scheduled to 2005, where it will 
ope\-rate for a 3 year period.
The instrument will be equipped with a proximity focusing RICH detector based on an aerogel 
radiator, enabling velocity measurements with a resolution of about 0.1\% and extending 
the charge measurements up to the Iron element.
The velocity of the cosmic rays is measured through the reconstruction of the 
$\check{C}$erenkov angle using a maximum Likelihood approach.
The method consists on finding the 
$\check{C}$erenkov angle maximising the overall pro\-ba\-bi\-li\-ty of the detected 
hits to belong to its cor\-res\-pon\- ding pattern.
Charge reconstruction is made in an event-by-event basis. 
It is based both on the velocity reconstruction procedure, which provides a reconstructed 
photon pattern, 
and on a semi-analytical calculation of the overall 
efficiency to detect the radiated $\check{C}$erenkov 
photons belonging to the reconstructed photon ring.

\end{document}